\documentclass[11pt,twoside]{article}

\usepackage{asp2006}
\usepackage{graphicx}
\usepackage{lscape}

\begin{document}
\title{The hard synchrotron X--ray spectrum of the TeV BL Lac 1ES 1426+428}
\author{A. Wolter\altaffilmark{1}, V. Beckmann\altaffilmark{2}, 
G. Ghisellini\altaffilmark{1}, F. Tavecchio\altaffilmark{1}, L. Maraschi\altaffilmark{1}, 
L. Costamante\altaffilmark{3}, A. Celotti\altaffilmark{4}, 
G. Ghirlanda\altaffilmark{1}}   
\altaffiltext{1}{INAF - Osservatorio Astronomico di Brera; Via Brera, 28 20121 MILANO, ITALY}
\altaffiltext{2}{INTEGRAL Science Data Centre, Ch.d'\`Ecogia 16, 1290 Versoix, Switzerland}
\altaffiltext{3}{Max-Planck Institut f\"ur Kernphysik, Saupfercheckweg 1, Heidelberg 69117, Germany}
\altaffiltext{4}{SISSA/ISAS, Via Beirut 2-4, I-34014 Trieste, Italy}
\begin{abstract} 
We have observed 1ES 1426+428 with INTEGRAL detecting it up to $\sim$150 keV. 
The spectrum is hard, confirming that this source is an extreme BL Lac object, 
with a synchrotron component peaking, in a $\nu F_\nu$ plot, at or above 100 keV,
resembling the hard states of Mkn 501 and 1ES 2344+514.
All these three sources are TeV emitters, with 1ES 1426+428 lying
at a larger redshift ($z=0.129$): for this source the absorption of
high energy photons by the IR cosmic background is particularly relevant.
The observed hard synchrotron tail helps the
modeling of its spectral energy distribution, giving information
on the expected intrinsic shape and flux in the TeV band.
This in turn constrains the amount of the poorly known IR background.
%
\end{abstract}


\section{Introduction}   

The spectral energy distribution (SED) of all blazars is characterized
by two broad peaks, in a $\nu F_\nu$ plot, located in the IR/X--ray
and in the MeV/TeV bands.  The origin of the low energy peak is almost
unanimously interpreted as synchrotron emission, while there is debate
about the origin of the high energy peak.  Coordinated and
simultaneous variability (Maraschi et al. 1999) strongly suggests that most of the
radiation of both components originates in one zone of the jet, and is
due to the same particles.  In the leptonic scenario the high energy
peak is inverse Compton emission by the same electrons making
synchrotron radiation at lower frequencies.  These electrons can
scatter the locally produced synchrotron photons (e.g., Maraschi et
al. 1992) and any other important component of seed photons produced
outside the jet, such as disk radiation (e.g., Dermer \& Schlickeiser
1993; Celotti et al. 2007), photons coming from the broad line region
(Sikora, Begelman \& Rees 1994; Ghisellini \& Madau 1996) or from the
layer of the jet in a spine/layer model (Ghisellini et al. 2005).

As suggested by Fossati et al. (1998) and Ghisellini et al. (1998) the
blazar SEDs seem to form a sequence: in powerful blazars the high
energy peak is dominant, and both peaks are at smaller frequencies.
As the bolometric luminosity decreases, blazars becomes ``bluer",
both peaks shifting to higher frequencies
(i.e. they become High Energy Peaked BL Lac, or HBL, in the
terminology of Padovani \& Giommi 1995), with approximately equal
power in the two peaks.  At the low luminosity end of this sequence we
then find BL Lacs which are TeV emitters.  The synchrotron component
of these BL Lacs peaks in the X--ray range, usually in the 0.1--1 keV
range, but a few extreme BL Lacs have their synchrotron peak (at least
occasionally), at or above $\sim$100 keV.

\begin{figure}[!htb]
\centerline{  \includegraphics[width=7cm,height=12cm,angle=-90]{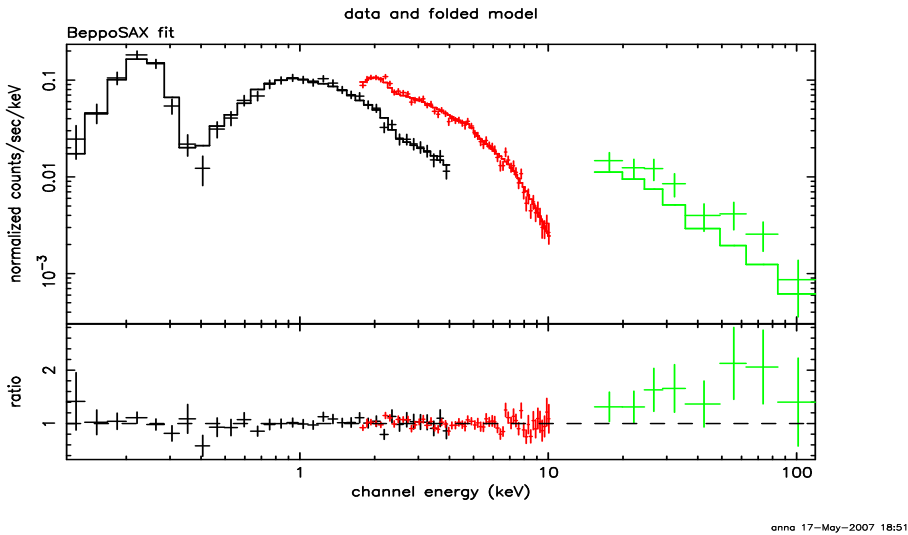}
}
\caption{
The BeppoSAX spectra of 1ES1426+428. 
The spectral fit is derived from the low
energy instruments, and extrapolated to the PDS range. The excess
at high energies is evident. A similar excess is seen in the
BeppoSAX observation of GB1428+4217. See Costamante et al. 2001 for details
of the fit.}
\label{sax}
\end{figure}

1ES 1426+428 was a candidate to belong to this extreme class: it was
observed by $Beppo$SAX (Costamante et al. 2001) with the narrow field
instruments (i.e. LECS, MECS between 0.1 and 10 keV) and the wide
field PDS instrument (betwen 20 and 100 keV).  The observed spectrum
was hard, up to 100 keV, but there was some uncertainty concerning the
PDS data, since this instrument could not resolve 1ES 1426+428 from
the nearby source (another blazar) GB 1428+4217.  Figure~\ref{sax} shows
the BeppoSAX X--ray spectrum in which the PDS excess is clearly
visible.  The INTEGRAL observations have
of course the required angular resolution to disentangle the two
sources, and have definitely confirmed the extreme nature of 1ES
1426+428.

The X-ray spectrum of 1ES 1426+428 resembles the hard states of Mkn 501 
and 1ES2344+514, all belonging to 
the class of extreme BL Lacs, and all detected in the TeV regime 
(eg. Albert et al. 2007, Smith et al. 2006). 
The knowledge of the high energy synchrotron spectrum is crucial to
construct a meaningful radiative model for the source, able to predict
the intrinsic spectrum in the TeV band, produced (within the leptonic
synchrotron self--Compton models) by the same electrons producing
X--rays by the synchrotron process.  This in turn is important to have
information about the amount of absorption the TeV flux have suffered,
due to the $\gamma$--$\gamma$ $\to$ $e^\pm$ process with the photons
of the IR cosmic background. The relatively high redshift of 
1ES 1426+428, z=0.129, makes it particularly interesting to study 
this effect.

\section{INTEGRAL data}   

A long look with INTEGRAL, 400 ksec, was given at this field in May 2006.
For a description of the INTEGRAL mission refer to Winkler et al. (2003).
The source was observed with the OMC (V filter), the JEM-X monitor 
(3--35 keV), the IBIS imager (0.02--10 MeV) and the SPI spectrometer
(0.02--8 MeV). The source was not detected by SPI
because it is below the detection limit. 

The blazar 1ES 1426+428 is detected, albeit at low 
significance, in both the JEM--X and the IBIS (ISGRI) instruments, while
the quasar GB 1428+4217 is not detected in any instrument.
We average the $V$--band light curve derived for 1ES 1426+428 from the OMC,
deriving m$_{V}$=16.0.

The SWIFT database ({\em http://heasarc.nasa.gov/docs/swift/archive/}) 
was \phantom{ } searched to find observations closer in time 
than BeppoSAX to the INTEGRAL one. One of the longest observations,
 taken on May 19th, 2005, was used to derive an X--ray spectrum 
for 1ES 1426+428. 

\begin{figure}[!htb]
\centerline{  \includegraphics[width=7cm,height=12cm,angle=-90]{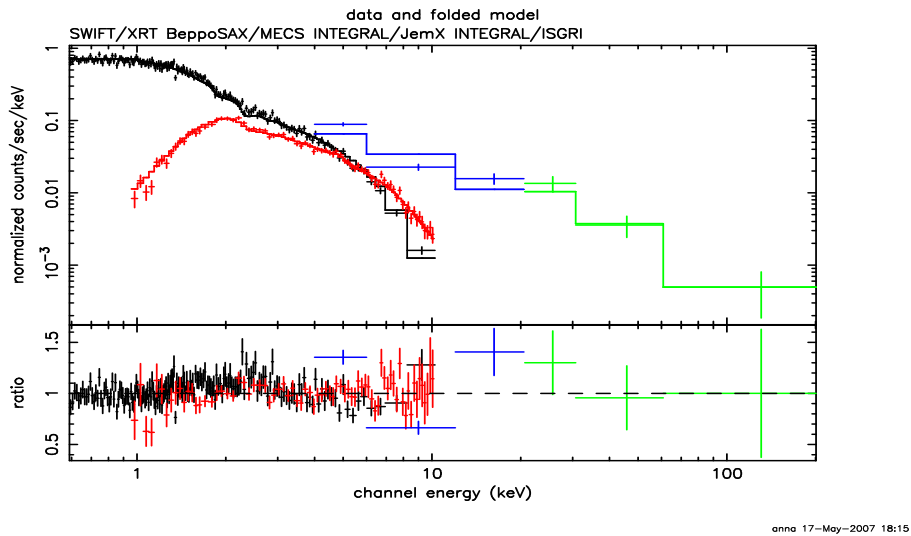}
}
\caption{
X--ray spectrum of 1ES 1426+428: from left to right the
datasets are Swift/XRT (black), BeppoSAX/MECS (red), INTEGRAL/JEM--X (blue)
and INTEGRAL/ISGRI (green).}
\label{fig3}
\end{figure}

Figure \ref{fig3} shows the resulting X--ray spectrum, combining the 
Swift/XRT, BeppoSAX/MECS, INTEGRAL/JEM-X and INTEGRAL/ISGRI datasets.\phantom{uff }
JEM-X is the most uncertain: it needs to be better calibrated.
In Table \ref{table} we report fluxes at different epochs, and a
combined spectral shape from the same instruments as above.

\begin{table*}[htb]
\caption{Table of fluxes and luminosities at different epochs with
the different instruments}
\begin{tabular}{llrrcc}
\hline
Instr. & Date & Flux {\tiny (2-10keV)}  & L$_X$ {\tiny (2-10keV)} &$\Gamma$ [1$\sigma$ range] & $\chi_r^2$(dof) \\
           &      & $(10^{-11}$ cgs) & $(10^{44}$ cgs)& & \\
\hline
XRT    & May 2005 &  3.16            & 13.9 & 1.98 [1.97--2.00] & 1.32(392)\\
MECS   & Feb 1999 &  2.01            & 8.8  &    &  \\
ISGRI  & May 2006 &  1.58/1.65$^{*}$ & 7.0  &    &  \\
\hline
\end{tabular}

{\tiny $^{*}$ the first number is the extrapolation to 2--10 keV,
the second is the measured flux in 20--100 keV.}
\label{table}
\end{table*}

\section{Discussion}   

\begin{figure}[!ht]
\centerline{ \includegraphics[width=11cm]{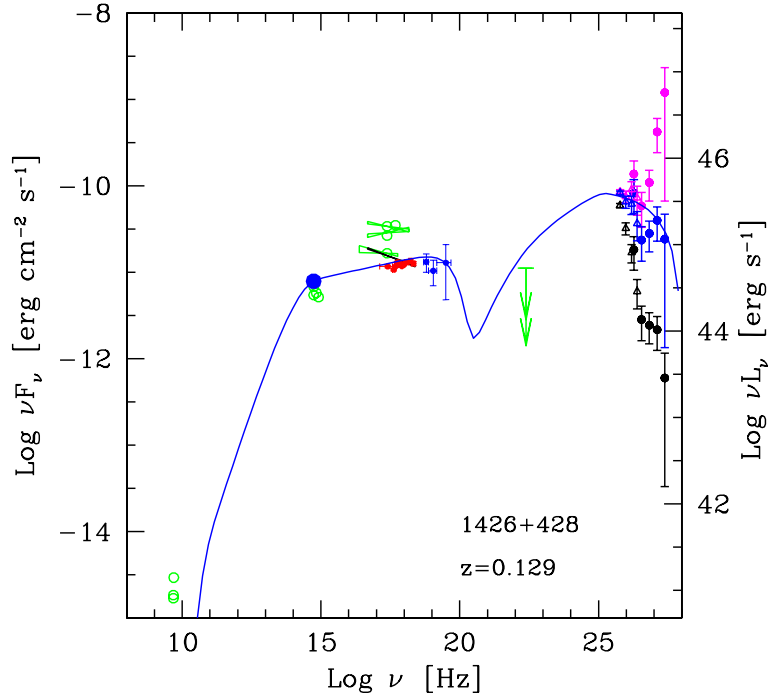}
}
\caption{The SED of 1ES 1426+428. Optical and ISGRI contemporaneous
data as blue circles; MECS spectrum in red and other X--ray data
(ASCA, ROSAT) in green; TeV observed points (CAT Djannati-Atai et
al. 2002, HEGRA, Aharonian et al. 2003) in black, with two different
de--absorption laws of the EBL (from Kneiske et al. (2004)): a milder
one (``low IR'') in blue, and a stronger one (``best'') in magenta.}
\label{sed}
\end{figure}

Figure~\ref{sed} shows the SED of 1ES 1426+428, using most of the
available datasets (see caption for details).  The correction for the
IR Extragalactic Background Light (IR EBL) is crucial for determining
the correct intrinsic shape. We report two deabsorbed spectra,
obtained using the ``low IR'' and ``best'' models of Kneiske et
al. (2004). The former case is compatible with the recent estimate of
the background of Aharonian et al. (2006).

The INTEGRAL spectrum shows that the peak of the synchrotron emission,
even if in a relatively low observed state, is at energies higher than
100 keV.  From ASCA data (Sambruna et al 1997) we know that the peak
was in the 2--10 keV range before 1999.  All Swift observations,
however, can be fitted (in the 2--10 keV range) with a power law with
$\Gamma \sim$ 2 (Wolter et al. in prep.), indicating that the spectral
shape did not change much recently, with flux variations of a factor
2--3.
Therefore we can reliably use these data to derive intrinsic quantities.

\subsection{Modeling the SED}

\begin{figure}[!ht]
\centerline{
 \includegraphics[width=12cm]{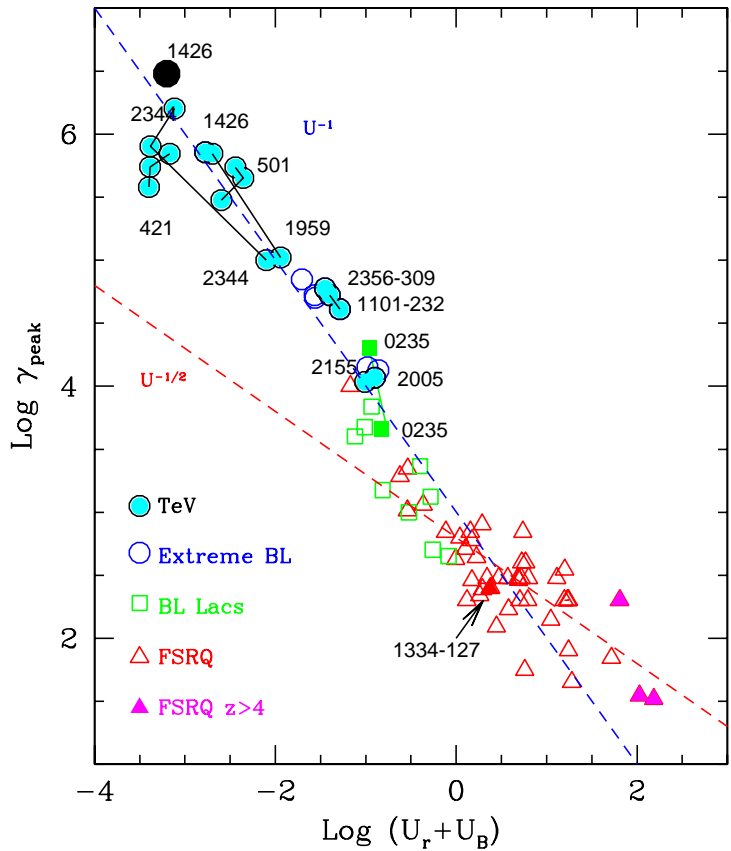}
}
\caption{
The random Lorentz factor $\gamma_{\rm peak}$ of the electrons emitting at the
peaks of the SED as a function of the energy density (magnetic plus
radiative, as seen in the comoving frame).
The black dot corresponds to the 1ES 1426+428
modelled in this paper (see Figure~\ref{sed}).
Adapted from Ghisellini et al. 2002.
}
\label{fig5}
\end{figure}

We have applied a simple, one--zone, homogeneous 
synchrotron self--Compton model, 
described in Ghisellini, Celotti \& Costamante (2002). 
The source is a cylinder of cross sectional radius
$R$ and width (in the comoving frame)
$\Delta R^\prime=R/\Gamma$, where $\Gamma$ is the bulk Lorentz
factor. 
The magnetic field $B$ is homogeneous and tangled.
The particle distribution is the result of injection and cooling.
We calculate the random Lorentz factor $\gamma_{\rm cool}$ at which
the particles cool in one light crossing time $\Delta R^\prime/c$.
Particles are injected between $\gamma_1$ and 
$\gamma_2$ with a power law distribution of slope $s$.
In our case  ($\gamma_{\rm cool}>\gamma_1$)
$N(\gamma) \propto \gamma^{-s}$ between $\gamma_1$ and 
$\gamma_{\rm cool}$, and
$N(\gamma) \propto \gamma^{-(s+1)}$ between $\gamma_{\rm cool}$ 
and $\gamma_2$.
Below $\min (\gamma_{\rm cool}, \gamma_1)$ we assume $N(\gamma)\propto \gamma^{-1}$.
We neglect any source of external photons.

Based on the above assumptions, the SED modeling yields the results shown
in Figure~\ref{sed}.
For the adopted parameters, we find that in one light crossing time
$\gamma_{\rm cool}>\gamma_2$, so that the particle distribution is
$N(\gamma)\propto \gamma^{-1}$ up to $\gamma_1=6000$, where it steepens
to $N(\gamma)\propto \gamma^{-2.86}$.
This electron slope makes a synchrotron spectrum $F(\nu)\propto \nu^{-0.93}$,
which is slightly increasing in a $\nu F_\nu$ plot.
As a consequence, the peak of the spectrum is made by the most energetic
electrons, $\gamma_{\rm peak} = \gamma_2= 3\times 10^6$.
The magnetic field is $B=0.08$ G, $R=8\times 10^{15}$ cm,
$\Gamma=18$, the viewing angle $\theta=1.8^\circ$, 
the Doppler factor $\delta=27.3$.
The power injected in relativistic electrons is $L^\prime=3\times 10^{41}$
erg s$^{-1}$.

These parameters are similar to the ones found
for other low power, TeV emitting, BL Lac objects (see e.g.
Katarzynski et al. 2006), even if
$\gamma_{\rm peak}$ is the largest ever found.

In Figure \ref{fig5} we show $\gamma_{\rm peak}$ as a function of the
comoving energy density (magnetic plus radiative) as seen in the
comoving frame, for a large sample of blazars.
Note that the values found here for 1ES 1426+428 are along the well defined
sequence, but reaching the extreme upper end in $\gamma_{\rm peak}$.

\section {Conclusions}

We observed 1ES 1426+428 with INTEGRAL: its spectrum is consistent 
with the $Beppo$SAX observation, with a small variation (40\%) in flux.
GB 1428+4217 is not detected by INTEGRAL, in any band.

A modeling of the SED in the standard SSC reference frame gives extreme
parameters (the largest $\gamma_{peak}$ ever found), and the source 
confirms the physical interpretation of the blazar sequence.
Our modeling prefers a moderate/low value of the cosmic background IR light.


%


\vskip 1truecm
\acknowledgements 

We  acknowledge partial financial support from the Italian
Space Agency (ASI) under contract ASI-INAF I/023/05/0.
This work is based on observations obtained with INTEGRAL, an ESA project
with instruments and science data centre funded by ESA member
states (especially the PI countries: Denmark, France, Germany,
Italy, Switzerland, Spain), Czech Republic and Poland, and with the
participation of Russia and the USA.


\end{document}